\documentclass{ifacconf}

\usepackage{graphicx}      
\usepackage{natbib}        

\usepackage{caption}
\usepackage{url}
\usepackage{amssymb,amsmath}
\usepackage[makeroom]{cancel}
\usepackage{epsfig}
\usepackage{epstopdf}
\usepackage{setspace}
\usepackage{enumerate}
\usepackage{tikz}
\usetikzlibrary{shapes,arrows}
\usepackage{algorithm}
\usepackage{algcompatible}
\usepackage{algpseudocode}
\usepackage{pifont}
\allowdisplaybreaks
\newtheorem{assumption}{Assumption}

\newtheorem{theorem}{Theorem}
\newtheorem{remark}{Remark}

\newtheorem{definition}{Definition}

\begin{document}
\begin{frontmatter}

\title{Distributed Nash Equilibrium Seeking via the Alternating Direction Method of Multipliers
} 


\author[First]{Farzad Salehisadaghiani} 
\author[First]{Lacra Pavel}

\address[First]{Department of Electrical and Computer Engineering, University of Toronto, Toronto, ON M5S 3G4, Canada (e-mails: {\tt\small farzad.salehisadaghiani@mail.utoronto.ca, pavel@ece.utoronto.ca}).}

\begin{abstract}                
In this paper, the problem of finding a Nash equilibrium of a multi-player game is considered. The players are only aware of their own cost functions as well as the action space of all players. We develop a relatively fast algorithm within the framework of inexact-ADMM. It requires a communication graph for the information exchange between the players as well as a few mild assumptions on cost functions. The convergence proof of the algorithm to a Nash equilibrium of the game is then provided. Moreover, the convergence rate is investigated via simulations.
\end{abstract}


\end{frontmatter}

\section{Introduction}
	There is a close connection between the problem of finding a Nash equilibrium (NE) of a distributed game and a distributed optimization problem. In a distributed optimization problem with $N$ agents that communicate over a connected graph, it is desired to minimize a global objective as follows:
\begin{equation}\label{mini}
\begin{cases}
\begin{aligned}
& \underset{x}{\text{minimize}}
& & f(x):=\sum_{i=1}^Nf_i(x) \\
& \text{subject to}
& & x\in \Omega.
\end{aligned}
\end{cases}
\end{equation}
In this problem the agents cooperatively solve \eqref{mini} over a common optimization variable $x$. In other words, all the agents are serving in the public interest in a way that they reduce the global loss. However, there are many real-world applications that involve selfishness of the players (agents) such as congestion control for Ad-hoc wireless networks and optical signal-to-noise ratio (OSNR) maximization in an optical network. In these applications, players \emph{selfishly} desire to optimize their own performance even though the global objective may not be minimized, hence play a game. In this regard, we are interested in studying the (Nash) equilibrium of this game.

Considering the difference between distributed optimization and distributed Nash equilibrium (NE) seeking, we aim to employ an optimization technique referred to as \emph{the alternating direction method of multipliers} (ADMM) to find an equilibrium point of a multi-player game. ADMM takes advantage of two different approaches used in solving optimization problems: 1) Dual Decomposition, and 2) Augmented Lagrangian Methods.  

Dual decomposition is a special case of a dual ascent method for solving an optimization problem when the objective function is separable w.r.t. variable $x$, i.e., $f(x):=\sum_{i=1}^Nf_i(x_i)$ where $x=[x_1,\ldots,x_N]^T$. This decomposition leads to $N$ parallel dual ascent problems whereby each is to be solved for $x_i$, $i\in\{1,\ldots,N\}$. This parallelism makes the convergence faster.

The augmented Lagrangian method is more robust and relaxes the assumptions in the dual ascent method. This method involves a penalty term added to the normal Lagrangian.  

In this work, we aim to exploit the benefits of ADMM in the context of finding an NE of a game. Here are the difficulties that we need to overcome:\vspace{-0.2cm}
\begin{itemize}
	\item A Nash game can be seen as a set of parallel optimization problems, each of them associated with the minimization of a player's own cost function w.r.t. his variable. However, each optimization problem is dependent on the solution of the other parallel problems. This leads to have $N$ Lagrangians whereby each is dependent on the other players' variables.  
	\item Each player $i$ updates only his own variable $x_i$, however, he requires also an estimate of all other variables $(x_j)_{j\in\{1,\ldots,N\},\,j\neq i}$ and updates it in order to solve his optimization problem. This demands an extra step in the algorithm based on communications between players.  
	\item Each optimization problem is not in the proper format of sum of separable functions to allow direct application of ADMM. 
\end{itemize} 

\emph{\textbf{Related Works.}} Our work is related to the literature on distributed Nash games such as \cite{yin2011nash,jayash8} and distributed optimization problems such as \cite{nedic2011asynchronous,johansson2008distributed}. Finding NE in distributed games has recently drawn attention due to many real-world applications. To name only a few, \cite{salehisadaghiani2014nash,salehisadaghiani2016distributed,frihauf2012nash,gharesifard2013distributed,salehisadaghiani2016arxiv,jayash10,jayash11}. In \cite{Jayash} an algorithm has been designed based on gossiping protocol to compute an NE in aggregative games. \cite{zhu2016distributed} study the problem of finding an NE in more general games by a gradient-based method over a complete communication graph. This problem is extended to the case with partially coupled cost functions (the functions which are not necessarily dependent on all the players' actions) in \cite{bramoulle2014strategic}. Recently, \cite{ye2015distributed} investigate distributed seeking of a time-varying NE with non-model-based costs for players. Computation of a time-varying NE is considered in \cite{lou2016nash} in networked games consisting of two subnetworks with shared objectives. \cite{parise2015network} propose two different algorithms to solve for an NE in a large population aggregative game which is subject to heterogeneous convex constraints.

ADMM algorithms, which are in the scope of this paper, have been developed in 1970s to find an optimal point of distributed optimization problems. This method has become widely used after its re-introduction in \cite{boyd2011distributed} such as \cite{he20121,goldstein2014fast,wei2012distributed}. \cite{shi2014linear} investigate the linear convergence rate of an ADMM algorithm to solve a distributed optimization problem. ADMM algorithms are extended by \cite{makhdoumi2014broadcast} to the case when agents broadcast their outcomes to their neighbors. The problem of distributed consensus optimization is considered in \cite{chang2015multi} which exploits \emph{inexact}-ADMM to reduce the computational costs of a classical ADMM. Recently, an ADMM-like algorithm is proposed by \cite{weipavelsubmitted} in order to find an NE of a game. It is shown that the algorithm converges faster than the gradient-based methods. However, the algorithm requires individual cocoercivity and is not developed, but rather postulated by mimicking of ADMM in distributed optimization according to the NE condition.  

\emph{\textbf{Contributions.}} In this paper, first, we reformulate the problem of finding an NE of a convex game as a set of distributed consensus optimization problems. Then we take advantage of a dummy variable to make the problem separable in the optimization variable. This technique can be used for any convex game which satisfies a set of relatively mild assumptions.

Second, we design a synchronous inexact-ADMM algorithm by which every player updates his action as well as his estimates of the other players' actions. This algorithm takes advantage of the speed and robustness of the classical ADMM and reduces the computational costs by using a linear approximation in players' action update rule (inexact-ADMM). Compared with gradient-based algorithms such as \cite{Jayash,zhu2016distributed,Marden2013}, our ADMM algorithm has an extra penalty term (could be seen as an extra state) which is updated through the iterations and improves the convergence rate. 

Third, we prove the convergence of the proposed algorithm toward the NE of the game and compare its convergence rate with a gradient-based method via simulation.

%
The paper is organized as follows. The problem statement and assumptions are provided in Section~2. In Section~3, an inexact-ADMM-like algorithm is proposed. Convergence of the algorithm to a Nash equilibrium of the game is discussed in Section~4 while in Section~5 a simplified representation of the algorithm for implementation is provided. Simulation results are given in Section~6 and conclusions in Section~7.\vspace{-0.2cm}
\section{Problem Statement}\label{problem_statement}

Consider $V=\{1,\ldots,N\}$ as a set of $N$ players in a networked multi-player game. The game is denoted by $\mathcal{G}$ and defined as follows:\vspace{-0.2cm}
\begin{itemize}
	\item $\Omega_i\subset\mathbb{R}$: Action set of player $i$, $\forall i\in V$,
	\item $\Omega=\prod_{i\in V}\Omega_i\subset\mathbb{R}^N$: Action set of all players,
	\item $J_i:\Omega\rightarrow \mathbb{R}$: Cost function of player $i$, $\forall i\in V$.
\end{itemize}

The game $\mathcal{G}(V,\Omega_i,J_i)$ is defined over the set of players, $V$, the action set of player $i\in V$, $\Omega_i$ and the cost function of player $i\in V$, $J_i$.

The players' actions are denoted as follows:\vspace{-0.2cm}
\begin{itemize}
	\item $x=(x_i,x_{-i})\in\Omega$: All players actions,
	\item $x_i\in\Omega_i$: Player $i$'s action, $\forall i\in V$,
	\item $x_{-i}\in\Omega_{-i}:=\prod_{j\in V\backslash\{i\}}\Omega_j$: All players' actions except player $i$'s.	
\end{itemize}
The game is played such that for a given $x_{-i}\in \Omega_{-i}$, each player $i\in V$ aims to minimize his own cost function selfishly w.r.t, $x_i$ to find an optimal action,
\begin{equation}
\label{mini_0}
\begin{cases}
\begin{aligned}
& \underset{x_i}{\text{minimize}}
& & J_i(x_i,x_{-i}) \\
& \text{subject to}
& & x_i\in \Omega_i.
\end{aligned}
\end{cases}
\end{equation}
Each optimization problem is run by a particular player $i$ at the same time with other players.

An NE of a game is defined as follows:
\begin{definition}\label{Nash_def}
	Consider an $N$-player game $\mathcal{G}(V,\Omega_i,J_i)$, each player $i$ minimizing the cost function $J_i:\Omega\rightarrow\mathbb{R}$. A vector $x^*=(x_i^*,x_{-i}^*)\in\Omega$ is called a Nash equilibrium of this game if
	\begin{equation}
	J_i(x_i^*,{x_{-i}^{*}})\leq J_i(x_{i},{x_{-i}^{*}})\quad\forall x_i\in \Omega_i,\,\,\forall i\in V.
	\end{equation}
\end{definition}

An NE lies at the intersection of all solutions of the set \eqref{mini_0}.
The challenge is that each optimization problem in \eqref{mini_0} is dependent on the solution of the other simultaneous problems. And since this game is distributed, no player is aware of the actions (solutions) of the other players (problems). 

We assume that each player $i$ maintains an estimate of the other players' actions. In the following, we define a few notations for players' estimates.\vspace{-0.2cm}
\begin{itemize}
	\item $x^i=(x_i^i,x_{-i}^i)\in\Omega$: Player $i$'s estimate of all players actions,
	\item $x_i^i\in\Omega_i$: Player $i$'s estimate of his own action, $\forall i\in V$,
	\item $x_{-i}^i\in\Omega_{-i}:=\prod_{j\in V\backslash\{i\}}\Omega_j$: Player $i$'s estimate of all other players' actions except his action,
	\item $\underline{x}=[{x^1}^T,\ldots,{x^N}^T]^T\in\Omega^N$: Augmented vector of estimates of all players' actions
\end{itemize} 
Note that player $i$'s estimate of his action is indeed his action, i.e., $x_i^i=x_i$ for $i\in V$. Note also that all players actions $x=(x_i,x_{-i})$ can be interchangeably represented as $x=[x_i^i]_{i\in V}$.

We assume that the cost function $J_i$ and the action set $\Omega$ are the only information available to player $i$. Thus, the players need to exchange some information in order to update their estimates. An undirected \emph{communication graph} $G_C(V,E)$ is defined where $E\subseteq V\times V$ denotes the set of communication links between the players. $(i,j)\in E$ if and only if players $i$ and $j$ exchange information. In the following, we have a few definitions for $G_C$:
\begin{itemize}\vspace{-0.2cm}
	\item $N_i:=\{j\in V|(i,j)\in E\}$: Set of neighbors of player $i$ in $G_C$,
	\item $A:=[a_{ij}]_{i,j\in V}$: Adjacency matrix of $G_C$ where $a_{ij}=1$ if $(i,j)\in E$ and $a_{ij}=0$ otherwise,
	\item $D:=\text{diag}\{|N_1|,\ldots,|N_N|\}$: Degree matrix of $G_C$. 
\end{itemize} 
The following assumption is used. 
\begin{assumption}\label{connectivity}
	$G_C$ is a connected graph.
\end{assumption}
We aim to relate game \eqref{mini_0} to the following problem whose solution can be based on the alternating direction method of multipliers (\cite{bertsekas1999parallel}, page 255).
\begin{equation*}
\begin{cases}
\begin{aligned}
& \underset{x\in C_1, z\in C_2}{\text{minimize}}
& & G_1(x)+G_2(z)\\
& \text{subject to}
& & Ax=z.
\end{aligned}
\end{cases}
\end{equation*}
To this end, we reformulate game \eqref{mini_0} so that the objective function is separable by employing estimates of the actions for each player $i\in V$ as $x^i$ (the estimates are also interpreted as the local copies of $x$). Particularly, from \eqref{mini_0}, consider that for a given $x_{-i}^i\in \Omega_{-i}$, each player $i\in V$ minimizes his cost function selfishly w.r.t. his own action subject to an equality constraint, i.e., for all $i\in V$,
\begin{equation*}
\label{mini_1-2}
\begin{cases}
\begin{aligned}
& \underset{x_i^i,x_i^j|_{j\in N_i}\in \Omega_i}{\text{minimize}}
& & J_i(x_i^i,x_{-i}^i)+\sum_{j\in N_i}g_j(x_i^j,x_{-i}^j)\\
& \hspace{0.25cm}\text{subject to}
& & x_i^i=x_i^j\quad\forall j\in N_i,
\end{aligned}
\end{cases}
\end{equation*}
where $g_j(\cdot)=0$ for $j\in N_i$. Note that, in order to update all elements in $x_{-i}^i$ we need to augment the constraint space to an $N\times 1$ vector form $x^i=x^j,\,j\in N_i$. Moreover, we replace the constraints with $x^l=x^s\quad\forall l\in V, \forall s\in N_l$ which includes $x^i=x^j,\,j\in N_i$. Note that augmenting the constraints in this way does not affect the solutions of the problem. Then for a given $x_{-i}^i\in \Omega_{-i}$ and for all $i\in V$, we obtain,
\begin{equation}
\label{mini_1}
\begin{cases}
\begin{aligned}
& \underset{x_i^i\in \Omega_i}{\text{minimize}}
& & J_i(x_i^i,x_{-i}^i)\\
& \text{subject to}
& & x^l=x^s\quad\forall l\in V, \forall s\in N_l.
\end{aligned}
\end{cases}
\end{equation}
The equality constraint along with Assumption~\ref{connectivity} ensures that all the local copies of $x$ are identical, i.e., $x^1=x^2=\ldots=x^N$. Hence \eqref{mini_1} recovers \eqref{mini_0}.

By Assumption~\ref{connectivity}, the set of problems \eqref{mini_1} are equivalent to the following set of optimization problems: for a given $x_{-i}^i\in\Omega_{-i}$ and for all $i\in V$,
\begin{equation}
\label{mini_2}
\begin{cases}
\begin{aligned}
& \underset{x_i^i\in \mathbb{R}}{\text{minimize}}
& & J_i(x_i^i,x_{-i}^i)+\mathcal{I}_{\Omega_i}(x_i^i)\\
& \text{subject to}
& & x^l=t^{ls}\quad\forall l\in V,\,\forall s\in N_l,\\
& & & x^s=t^{ls}\quad\forall l\in V,\,\forall s\in N_l,
\end{aligned}
\end{cases}
\end{equation}
where $\mathcal{I}_{\Omega_i}(x_i^i):=\begin{cases}0&\text{if }x_i^i\in\Omega_i\\\infty&\text{otherwise}\end{cases}$ is an indicator function of the feasibility constraint $x_i^i\in\Omega_i$ and $t^{ls}$ is an intermediary variable to separate the equality constraints.

Note that one can regard the set of problems \eqref{mini_2} as being the same as the set of problems \eqref{mini_0} but considering $N$ estimates (local copies) of the players' actions for each player $i\in V$.

A characterization of the NE for game \eqref{mini_0} could be obtained by finding KKT conditions on the set of problems~\eqref{mini_2}. Let $\{u^{ls},v^{ls}\}_{l\in V,s\in N_l}$ with $u^{ls},v^{ls}\in \mathbb{R}^N$ be the Lagrange multipliers associated with the two constraints in \eqref{mini_2}, respectively. The corresponding Lagrange function for player $i$, $\forall i\in V$ is as follows:\vspace{-0.3cm}
\begin{eqnarray}
&&\hspace{-1.7cm}L_i\Big(x^i, \{t^{ls}\}_{l\in V,s\in N_l}, \{u^{ls}\}_{l\in V,s\in N_l}, \{v^{ls}\}_{l\in V,s\in N_l}\Big)\nonumber\\
&&\hspace{-1.7cm}:= J_i(x_i^i,x_{-i}^i)+\mathcal{I}_{\Omega_i}(x_i^i)+\sum_{l\in V}\sum_{s\in N_l}{u^{ls}}^T(x^l-t^{ls})\nonumber\\
&&\hspace{-1.7cm}+\sum_{l\in V}\sum_{s\in N_l}{v^{ls}}^T(x^s-t^{ls}),
\end{eqnarray}
Let $({x^i}^*)_{i\in V}$ and $\{{u^{ls}}^*,{v^{ls}}^*\}_{l\in V,\,s\in N_l}$ be a pair of optimal primal and dual solutions to \eqref{mini_2}. The KKT conditions are summarized as follows:\vspace{-0.2cm}
\begin{eqnarray}
&&\hspace{-1cm}\nabla_iJ_i({x^i}^*)+\partial_i\mathcal{I}_{\Omega_i}({x_i^i}^*)+\sum_{j\in N_i}{u_i^{ij}}^*+{v_i^{ji}}^*=0\,\,\,\,\forall i\in V,\label{PNash1}\\
&&\hspace{-1cm}{x^i}^*={x^j}^*\quad\forall i\in V,\,\forall j\in N_i,\label{PNash2}\\
&&\hspace{-1cm}{u^{ij}}^*+{v^{ij}}^*=\textbf{0}_N\quad\forall i\in V,\,\forall j\in N_i,\label{PNash3}
\end{eqnarray}
where $\nabla_iJ_i(\cdot)$ is gradient of $J_i$ w.r.t. $x_i$ and $\partial_i\mathcal{I}_{\Omega_i}(\cdot)$ is a subgradient of $\mathcal{I}_{\Omega_i}$ at $x_i$. Note that the index of ${v^{ji}}^*$ in \eqref{PNash1} is inverse of ${v^{ij}}^*$ in \eqref{PNash3}. By \eqref{PNash2} and Assumption~\ref{connectivity}, ${x^1}^*=\ldots={x^N}^*:=x^*$. Thus, $x^*:=(x_i^*,x_{-i}^*)$ is a solution of \eqref{mini_2} if and only if,\vspace{-0.3cm}
\begin{eqnarray}\label{Nash_equations}
\begin{cases}
\nabla_iJ_i({x}^*)+\partial\mathcal{I}_{\Omega_i}({x_i}^*)+\sum_{j\in N_i}{u_i^{ij}}^*+{v_i^{ji}}^*=0\,\,\,\,\forall i\in V,\\
{u^{ij}}^*+{v^{ij}}^*=\textbf{0}_N\quad\forall i\in V,\,\forall j\in N_i.
\end{cases}
\end{eqnarray}	 

We state a few assumptions for the existence and the uniqueness of an NE.
\begin{assumption}
	\label{assump}
	For every $i\in V$, the action set $\Omega_i$ is a non-empty, compact and convex subset of $\mathbb{R}$. $J_i(x_i,x_{-i})$ is a continuously differentiable function in $x_i$, jointly continuous in $x$ and convex in $x_i$ for every $x_{-i}$.
\end{assumption}
The convexity of $\Omega_i$ implies that $\mathcal{I}_{\Omega_i}$ is a convex function. This yields that there exists at least one bounded subgradient $\partial\mathcal{I}_{\Omega_i}$. 
\begin{assumption}\label{Lip_assump}
	Let $F:\Omega^N\rightarrow\mathbb{R}^N$, $F(\underline{x}):=[\nabla_iJ_i(x^i)]_{i\in V}$ be the the pseudo-gradient vector (game map) where $\underline{x}:=[{x^1}^T,\ldots,{x^N}^T]^T\in\Omega^N$. $F$ is cocoercive $\forall\underline{x}\in\Omega^N$ and $\underline{y}\in\Omega^N$, i.e.,\vspace{-0.4cm}
	\begin{eqnarray}
	&&\hspace{-1cm}(F(\underline{x})-F(\underline{y}))^T(x-y)\geq \sigma_F\|F(\underline{x})-F(\underline{y})\|^2,
	\end{eqnarray}
	where $\sigma_F>0$.
\end{assumption}
\begin{remark}
Assumption~\ref{assump} is a standard assumption in the literature of NE seeking. Assumption~\ref{Lip_assump} is relatively stronger than the (strong) monotonicity of the game map (pseudo-gradient vector) (see \cite{zhu2016distributed,Jayash}). However, as we will show, this leads to an algorithm with the benefits of ADMM algorithms (speed).
\end{remark}
\begin{remark}
Assumption~\ref{Lip_assump} is not usually required in distributed optimization problems; there instead the (strong) convexity of the objective function is assumed to be w.r.t. the full vector $x$ and also the gradient of the objective function is assumed to be Lipschitz continuous (see \cite{chang2015multi}).
\end{remark}
Our objective is to find an ADMM-like\footnote[1]{ADMM or Alternative Direction Method of Multipliers also known as Douglas-Rachford splitting is a method of solving an optimization problem where the objective function is a summation of two convex (possibly non-smooth) functions. For a detailed explanation see \cite{parikh2014proximal}.} algorithm for computing an NE of $\mathcal{G}(V,\Omega_i,J_i)$ using only imperfect information over the communication graph $G_C(V,E)$.\vspace{-0.2cm}
\section{Distributed Inexact ADMM Algorithm}\label{asynch}
We propose a distributed algorithm, using an \emph{inexact\footnote[2]{In an inexact consensus ADMM instead of solving an optimization sub-problem, a method of approximation is employed to reduce the complexity of the sub-problem.} consensus ADMM}. We obtain an NE of $\mathcal{G}(V,\Omega_i,J_i)$ by solving the set of problems \eqref{mini_2} by an ADMM-like approach.

The mechanism of the algorithm can be briefly explained as follows:
Each player maintains an estimate of the actions of all players and locally communicates with his neighbors over $G_C$. Then, he takes average of his neighbors' information and uses it to update his estimates.

The algorithm is elaborated in the following steps:\\
1- \textbf{\emph{Initialization Step:}}
Each player $i\in V$ maintains an initial estimate for all players, $x^i(0)\in\Omega$. The initial values of $u^{ij}(0)$ and $v^{ij}(0)$ are set to be zero for all $i\in V$, $j\in N_i$.\\
2- \textbf{\emph{Communication Step:}}
At iteration $T(k)$, each player $i\in V$ exchanges his estimate of the other players' actions with his neighbors $j$, $\forall j\in N_i$. Then, he takes average of the received information with his estimate and updates his estimate as follows:\vspace{-0.2cm}
\begin{eqnarray}
&&\hspace{-1cm}x_{-i}^i(k)=\frac{1}{2}\Big(x_{-i}^i(k-1)+\underbrace{\frac{1}{|N_i|}\sum_{j\in N_i}x_{-i}^j(k-1)}_{\text{RECEIVED INFORMATION}}\Big)\nonumber\\
&&\hspace{-1cm}-\underbrace{\frac{1}{2c|N_i|}\sum_{j\in N_i}(u_{-i}^{ij}(k)+v_{-i}^{ji}(k))}_{\text{PENALTY TERM}},\label{x_-i_Update}
\end{eqnarray}
where $c>0$ is a scalar coefficient, and $\forall i\in V,j\in N_i$,\vspace{-0.2cm}
\begin{eqnarray}
&&\hspace{-1cm}u^{ij}(k)=u^{ij}(k-1)+\frac{c}{2}\big(x^i(k-1)-x^j(k-1)\big),\label{u_update_ADMM}\\
&&\hspace{-1cm}v^{ij}(k)=v^{ij}(k-1)+\frac{c}{2}\big(x^j(k-1)-x^i(k-1)\big).\label{v_update_ADMM}	\end{eqnarray}
Equations \eqref{u_update_ADMM}, \eqref{v_update_ADMM} are the dual Lagrange multipliers update rules. Note that in \eqref{x_-i_Update}, a penalty factor $\sum_{j\in N_i}(u_{-i}^{ij}(k)+v_{-i}^{ji}(k))$ is subtracted, which is associated with the difference between the estimates of the neighboring players (Equations~\eqref{u_update_ADMM}, \eqref{v_update_ADMM}).
\begin{remark}\label{vs_opt}
Unlike distributed optimization algorithms where the minimization is w.r.t. $x$, here each player minimizes his cost function w.r.t. $x_i^i$. To update $x_i^i$, each player requires the estimate of the other players $x_{-i}^i$ at each iteration. Thus, the communication step is inevitable to update $x_{-i}^i$ for the next iteration. 
\end{remark}
3- \textbf{\emph{Action Update Step}}

At this moment all the players update their actions via an ADMM-like approach developed as follows. For each player $i$, $\forall i\in V$ let the augmented Lagrange function associated to problem \eqref{mini_2} be as follows:
\begin{eqnarray}\label{augmented_Lag}
&&\hspace{-1.7cm}L_i^{a}\Big(x^i, \{t^{ls}\}_{l\in V,s\in N_l}, \{u^{ls}\}_{l\in V,s\in N_l}, \{v^{ls}\}_{l\in V,s\in N_l}\Big)\nonumber\\
&&\hspace{-1.7cm}:= J_i(x_i^i,x_{-i}^i)+\mathcal{I}_{\Omega_i}(x_i^i)\nonumber\\
&&\hspace{-1.7cm}+\sum_{l\in V}\sum_{s\in N_l}{u^{ls}}^T(x^l-t^{ls})+\sum_{l\in V}\sum_{s\in N_l}{v^{ls}}^T(x^s-t^{ls})\nonumber\\
&&\hspace{-1.7cm}+\frac{c}{2}\sum_{l\in V}\sum_{s\in N_l}(\|x^l-t^{ls}\|^2+\|x^s-t^{ls}\|^2).
\end{eqnarray}
where $c>0$ is a scalar coefficient which is also used in \eqref{x_-i_Update}, \eqref{u_update_ADMM} and \eqref{v_update_ADMM}. Consider the ADMM algorithm associated to problem \eqref{mini_2} based on \eqref{augmented_Lag}:\vspace{-0.3cm}
\begin{eqnarray}
&&\hspace{-0.8cm}x_i^i(k)\!=\!\text{arg }\min_{x_i^i\in\mathbb{R}}\!L_i^a\Big(\!(x_i^i,x_{-i}^i(k-1))\!,\!\{t^{ls}(k-1)\}_{l\in V,s\in N_l}\!,\nonumber\\
&&\hspace{-0.8cm}\{u^{ls}(k)\}_{l\in V,s\in N_l}, \{v^{ls}(k)\}_{l\in V,s\in N_l}\Big)\nonumber\\
&&\hspace{-0.8cm}=\text{arg }\min_{x_i^i\in\mathbb{R}}\Big\{J_i(x_i^i,x_{-i}^i(k-1))+\mathcal{I}_{\Omega_i}(x_i^i)\nonumber\\
&&\hspace{-0.8cm}+\sum_{j\in N_i}(u^{ij}(k)+v^{ji}(k))^T(x_i^i,x_{-i}^i(k-1))\nonumber\\
&&\hspace{-0.8cm}+c\sum_{j\in N_i}\Big\|(x_i^i,x_{-i}^i(k-1))-t^{ij}(k-1)\Big\|^2\Big\}\quad\forall i\in V,\label{x_i^i_ADMM_bef}
\end{eqnarray}
The update rule for the auxiliary variable $t^{ij}$ $\forall i\in V, j\in N_i$ is based on \eqref{augmented_Lag},\vspace{-0.3cm}
\begin{eqnarray}
&&\hspace{-1cm}t^{ij}(k)=\text{arg }\min_{t^{ij}} L_i^a\Big((x_i^i(k),x_{-i}^i(k-1)), \{t^{ls}\}_{l\in V,s\in N_l},\nonumber\\
&&\hspace{-1cm}\{u^{ls}(k)\}_{l\in V,s\in N_l}, \{v^{ls}(k)\}_{l\in V,s\in N_l}\Big)\nonumber\\
&&\hspace{-1cm}=\text{arg }\min_{t^{ij}}\Big\{-(u^{ij}(k)+v^{ij}(k))^Tt^{ij}\nonumber\\
&&\hspace{-1cm}+\frac{c}{2}(\|(x_i^i(k),x_{-i}^i(k-1))-t^{ij}\|^2\nonumber\\
&&\hspace{-1cm}+\|(x_i^j(k),x_{-i}^j(k-1))-t^{ij}\|^2)\Big\}=\frac{1}{2c}(u^{ij}(k)+v^{ij}(k))\nonumber\\
&&\hspace{-1cm}+\frac{1}{2}((x_i^i(k),x_{-i}^i(k-1))+(x_i^j(k),x_{-i}^j(k-1))).\label{x_i^i_ADMM}
\end{eqnarray}
The initial conditions $u^{ij}(0)=v^{ij}(0)=\textbf{0}_N$ $\forall i\in V,\,j\in N_i$ along with \eqref{u_update_ADMM} and \eqref{v_update_ADMM} suggest that $u^{ij}(k)+v^{ij}(k)=\textbf{0}_N$ $\forall i\in V,\,j\in N_i,\,k>0$. Then,
\begin{equation}
t^{ij}(k)=\frac{(x_i^i(k),x_{-i}^i(k-1))+(x_i^j(k),x_{-i}^j(k-1))}{2}.\label{t_ij_simp}
\end{equation}
Using \eqref{t_ij_simp} in \eqref{x_i^i_ADMM_bef}, one can derive the local estimate update for all $i\in V$ as follows:\vspace{-0.3cm}
\begin{eqnarray}\label{x_i^i_ADMM_Pre}
&&\hspace{-0.8cm}x_i^i(k)=\text{arg }\min_{x_i^i\in\mathbb{R}}\Big\{J_i(x_i^i,x_{-i}^i(k-1))+\mathcal{I}_{\Omega_i}(x_i^i)\nonumber\\
&&\hspace{-0.8cm}+\sum_{j\in N_i}(u^{ij}(k)+v^{ji}(k))^T(x_i^i,x_{-i}^i(k-1))\nonumber\\		&&\hspace{-0.8cm}+c\sum_{j\in N_i}\Big\|(x_i^i,x_{-i}^i(k-1))\\
&&\hspace{-0.8cm}-\frac{(x_i^i(k-1),x_{-i}^i(k-2))+(x_i^j(k-1),x_{-i}^j(k-2))}{2}\Big\|^2\Big\}\nonumber
\end{eqnarray}
We simplify \eqref{x_i^i_ADMM_Pre} by using a proximal first-order approximation for $J_i(x_i^i,x_{-i}^i(k-1))$ around $x^i(k-1)$; thus  using inexact ADMM it follows:\vspace{-0.3cm}
\begin{eqnarray}
&&\hspace{-0.2cm}x_i^i(k)=\text{arg }\min_{x_i^i\in\mathbb{R}}\Big\{\nabla_iJ_i(x^i(k-1))^T(x_i^i-x_{i}^i(k-1))\nonumber\\
&&\hspace{-0.2cm}+\frac{\beta_i}{2}\|x_i^i-x_{i}^i(k-1)\|^2+\mathcal{I}_{\Omega_i}(x_i^i)+\sum_{j\in N_i}(u_{i}^{ij}(k)+v_{i}^{ji}(k))x_i^i\nonumber\\
&&\hspace{-0.2cm}+c\sum_{j\in N_i}\Big\|x_i^i-\frac{x_i^i(k-1)+x_i^j(k-1)}{2}\Big\|^2\Big\}\quad\forall i\in V,\label{x_i^i_Mod_ADMM_1}
\end{eqnarray}
where $\beta_i>0$ is a penalty factor for the proximal first-order approximation of each player $i$'s cost function. 

At this point, the players are ready to begin a new iteration from step 2. To sum up, the algorithm consists of \eqref{x_-i_Update}, \eqref{u_update_ADMM}, \eqref{v_update_ADMM} and \eqref{x_i^i_Mod_ADMM_1} which are the update rule for the players' estimates except their own actions, the update rules for the Lagrange multipliers and the update rule for player's action, respectively.\vspace{-0.2cm}
\section{Convergence Proof}\label{convergence_proof}
\begin{theorem}\label{theorem_convergence_rate}
	Let $\beta_{\min}:=\min_{i\in V}\beta_i>0$\footnote[3]{In order to have a fully distributed algorithm, one can consider a network-wide known lower bound $\tilde{\beta}_{\min}$, $\tilde{\beta}_{\min}\leq\beta_i\,\forall i\in V$ and use it instead of $\beta_{\min}$.}be the minimum penalty factor of the approximation in the inexact ADMM algorithm which satisfies\vspace{-0.2cm}
	\begin{equation}\label{condition}
	\sigma_F>\frac{1}{2(\beta_{\min}+c\lambda_{\min}(D+A))},
	\end{equation}
	where $\sigma_F$ is a positive constant for the cocoercive property of $F$, and $D$ and $A$ are the degree and adjacency matrices of $G_C$, respectively. Under Assumptions~\ref{connectivity}-\ref{Lip_assump}, the sequence $\{x^i(k)\}$ $\forall i\in V$, generated by the algorithm \eqref{x_-i_Update}, \eqref{u_update_ADMM}, \eqref{v_update_ADMM} and \eqref{x_i^i_Mod_ADMM_1}, converges to $x^*$ NE of game \eqref{mini_0}. 
\end{theorem}
\par{\emph{Proof}}.
The optimality condition of \eqref{x_i^i_Mod_ADMM_1} yields:\vspace{-0.2cm}
\begin{eqnarray}\label{optim_equation}
&&\hspace{-0.8cm}\nabla_iJ_i(x^i(k-1))+\beta_i(x_i^i(k)-x_{i}^i(k-1))\nonumber\\
&&\hspace{-0.8cm}+\partial_i\mathcal{I}_{\Omega_i}(x_i^i(k))+\sum_{j\in N_i}(u_i^{ij}(k)+v_i^{ji}(k))\nonumber\\
&&\hspace{-0.8cm}+2c\sum_{j\in N_i}\Big(x_i^i(k)-\frac{x_i^i(k-1)+x_i^j(k-1)}{2}\Big)=0.
\end{eqnarray}	
We combine \eqref{optim_equation} with \eqref{Nash_equations} which represents the equations associated with the solutions of the set of problems \eqref{mini_2} (NE of game \eqref{mini_0}). Then we obtain,\vspace{-0.2cm}
\begin{eqnarray}\label{optim_equation_Nash_combo}
&&\hspace{-1cm}\nabla_iJ_i(x^i(k-1))-\nabla_iJ_i(x^*)+\beta_i(x_i^i(k)-x_{i}^i(k-1))\nonumber\\
&&\hspace{-1cm}+\partial\mathcal{I}_{\Omega_i}(x_i^i(k))-\partial\mathcal{I}_{\Omega_i}(x_i^*)\nonumber\\
&&\hspace{-1cm}+\sum_{j\in N_i}(u_i^{ij}(k)+v_i^{ji}(k)-{u_i^{ij}}^*-{v_i^{ji}}^*)\nonumber\\
&&\hspace{-1cm}+2c\sum_{j\in N_i}\Big(x_i^i(k)-\frac{x_i^i(k-1)+x_i^j(k-1)}{2}\Big)=0.
\end{eqnarray}
We multiply both sides by $(x_i^i(k)-x_i^*)$ and then add and subtract $x_i^i(k-1)$ as follows:\vspace{-0.3cm}
\begin{eqnarray}\label{optim_equation_Nash_combo_multiplic}
&&\hspace{-0.8cm}\Big(\nabla_iJ_i(x^i(k-1))-\nabla_iJ_i(x^*)\Big)^T(x_i^i(k-1)-x_i^*)\nonumber\\
&&\hspace{-0.8cm}+\Big(\nabla_iJ_i(x^i(k-1))-\nabla_iJ_i(x^*)\Big)^T(x_i^i(k)-x_i^i(k-1))\nonumber\\
&&\hspace{-0.8cm}+\beta_i(x_i^i(k)-x_{i}^i(k-1))^T(x_i^i(k)-x_i^*)\nonumber\\
&&\hspace{-0.8cm}+(\partial\mathcal{I}_{\Omega_i}(x_i^i(k))-\partial\mathcal{I}_{\Omega_i}(x_i^*))^T(x_i^i(k)-x_i^*)\nonumber\\
&&\hspace{-0.8cm}+\sum_{j\in N_i}(u_i^{ij}(k)+v_i^{ji}(k)-{u_i^{ij}}^*-{v_i^{ji}}^*)^T(x_i^i(k)-x_i^*)\\
&&\hspace{-0.8cm}+2c\!\sum_{j\in N_i}\!\Big(\!x_i^i(k)\!-\!\frac{x_i^i(k-1)\!+\!x_i^j(k-1)}{2}\!\Big)^T\!(x_i^i(k)\!-\!x_i^*)\!=\!0.\nonumber
\end{eqnarray}	
As discussed in Remark~\ref{vs_opt}, in addition to updating their own actions, the players need to update their estimates as well. In the following, we explain how to bring in the update rule of $x_{-i}^i$ into \eqref{optim_equation_Nash_combo_multiplic}.    

Note that by \eqref{x_-i_Update}, one can obtain,\vspace{-0.3cm}
\begin{eqnarray}\label{u_for_-i}
&&\hspace{-0.6cm}\sum_{j\in N_i}(u_{-i}^{ij}(k)+v_{-i}^{ji}(k))\\
&&\hspace{-0.6cm}+2c\sum_{j\in N_i}\Big(x_{-i}^i(k)-\frac{x_{-i}^i(k-1)+x_{-i}^j(k-1)}{2}\Big)=\textbf{0}_{N-1}.\nonumber
\end{eqnarray}
Multiplying \eqref{u_for_-i} by $(x_{-i}^i(k)-x_{-i}^*)$, one can arrive at,
\begin{eqnarray}\label{u_for_-i_*_mult}
&&\hspace{-1cm}\sum_{j\in N_i}(u_{-i}^{ij}(k)+v_{-i}^{ji}(k))^T(x_{-i}^i(k)-x_{-i}^*)\nonumber\\
&&\hspace{-1cm}+2c\sum_{j\in N_i}\Big(x_{-i}^i(k)-\frac{x_{-i}^i(k-1)+x_{-i}^j(k-1)}{2}\Big)^T\nonumber\\
&&\hspace{-1cm}.(x_{-i}^i(k)-x_{-i}^*)=0.
\end{eqnarray}
Adding \eqref{u_for_-i_*_mult} to \eqref{optim_equation_Nash_combo_multiplic} and using \eqref{u_update_ADMM}, \eqref{v_update_ADMM}, yilds $\forall i\in V$,\vspace{-0.3cm}
\begin{eqnarray}\label{optim_equation_Nash_combo_multiplic_Revised}
&&\hspace{-1cm}\Big(\nabla_iJ_i(x^i(k-1))-\nabla_iJ_i(x^*)\Big)^T(x_i^i(k-1)-x_i^*)\nonumber\\
&&\hspace{-1cm}+\Big(\nabla_iJ_i(x^i(k-1))-\nabla_iJ_i(x^*)\Big)^T(x_i^i(k)-x_i^i(k-1))\nonumber\\
&&\hspace{-1cm}+\beta_i(x_i^i(k)-x_{i}^i(k-1))^T(x_i^i(k)-x_i^*)\nonumber\\
&&\hspace{-1cm}+(\partial\mathcal{I}_{\Omega_i}(x_i^i(k))-\partial\mathcal{I}_{\Omega_i}(x_i^*))^T(x_i^i(k)-x_i^*)\nonumber\\
&&\hspace{-1cm}+\!\sum_{j\in N_i}\!(\!u_i^{ij}(k+1)\!+\!v_i^{ji}(k+1)\!-\!{u_i^{ij}}^*\!-\!{v_i^{ji}}^*\!)^T\!(x_i^i(k)\!-\!x_i^*)\nonumber\\
&&\hspace{-1cm}+\sum_{j\in N_i}(u_{-i}^{ij}(k+1)+v_{-i}^{ji}(k+1))^T(x_{-i}^i(k)-x_{-i}^*)\nonumber\\
&&\hspace{-1cm}+2c\sum_{j\in N_i}\Big(\frac{x^i(k)+x^j(k)}{2}-\frac{x^i(k-1)+x^j(k-1)}{2}\Big)^T\nonumber\\
&&\hspace{-1cm}.(x^i(k)-x^*)=0.
\end{eqnarray}
The second and the third terms are bounded as follows:\vspace{-0.3cm}
\begin{eqnarray}\label{cauchy_Schewwrtz}
&&\hspace{-0.6cm}\Big(\nabla_iJ_i(x^i(k-1))-\nabla_iJ_i(x^*)\Big)^T(x_i^i(k)-x_i^i(k-1))\geq\\
&&\hspace{-0.6cm}\frac{-1}{2\rho}\!\|\!\nabla_iJ_i(x^i(k-1))\!-\!\nabla_iJ_i(x^*)\!\|^2\!-\!\frac{\rho}{2}\!\|\!x_i^i(k)\!-\!x_i^i(k-1)\!\|^2,\nonumber
\end{eqnarray}
for any $\rho>0$ $\forall i\in V$. By the convexity of $\mathcal{I}_{\Omega_i}$ (Assumption~\ref{assump}), we have for the fourth term,
\begin{equation}\label{convexity_of_Indicator}
(\partial\mathcal{I}_{\Omega_i}(x_i^i(k))-\partial\mathcal{I}_{\Omega_i}(x_i^*))^T(x_i^i(k)-x_i^*)\geq 0.
\end{equation}
Using \eqref{cauchy_Schewwrtz} and \eqref{convexity_of_Indicator} in \eqref{optim_equation_Nash_combo_multiplic_Revised} and summing over $i\in V$, we obtain,\vspace{-0.3cm}
\begin{eqnarray}\label{optim_equation_Nash_combo_multiplic_Revised_Sum}
&&\hspace{-0.7cm}\Big(F(\underline{x}(k-1))-F(\underline{x}^*)\Big)^T(x(k-1)-x^*)\nonumber\\
&&\hspace{-0.7cm}-\frac{1}{2\rho}\|F(\underline{x}(k-1))-F(\underline{x}^*)\|^2-\frac{1}{2}\|\underline{x}(k)-\underline{x}(k-1)\|_{M_1}^2\nonumber\\
&&\hspace{-0.7cm}+(\underline{x}(k)-\underline{x}(k-1))^T\text{diag}((\beta_ie_ie_i^T)_{i\in V})(\underline{x}(k)-\underline{x}^*)\nonumber\\
&&\hspace{-0.7cm}+\!\sum_{i\in V}\!\sum_{j\in N_i}\!(\!u_i^{ij}(k+1)\!+\!v_i^{ji}(k+1)\!-\!{u_i^{ij}}^*\!-\!{v_i^{ji}}^*\!)^T\!(\!x_i^i(k)\!-\!x_i^*\!)\nonumber\\
&&\hspace{-0.7cm}+\sum_{i\in V}\sum_{j\in N_i}(u_{-i}^{ij}(k+1)+v_{-i}^{ji}(k+1))^T(x_{-i}^i(k)-x_{-i}^*)\nonumber\\
&&\hspace{-0.7cm}+c(\underline{x}(k)-\underline{x}(k-1))^T((D+A)\otimes I_N)(\underline{x}(k)-\underline{x}^*)\leq 0,
\end{eqnarray}
where $M_1:=\text{diag}((\rho e_ie_i^T)_{i\in V})$ and $\underline{x}^*=[{{x^1}^*}^T,\ldots,{{x^N}^*}^T]^T$. We bound the first term using Assumption~\ref{Lip_assump},\vspace{-0.3cm}
\begin{eqnarray}\label{Cocoe_in_equat}
&&\hspace{-1cm}\Big(F(\underline{x}(k-1))-F(\underline{x}^*)\Big)^T(x(k-1)-x^*)\nonumber\\
&&\hspace{-1cm}\geq\sigma_F\|F(\underline{x}(k-1))-F(\underline{x}^*)\|^2.
\end{eqnarray}
We also simplify the fifth and the sixth terms in \eqref{optim_equation_Nash_combo_multiplic_Revised_Sum}. Since $G_C$ is an undirected graph, for any $\{a_{ij}\}$, $\sum_{i\in V}\sum_{j\in N_i}a_{ij}=\sum_{i\in V}\sum_{j\in N_i}a_{ji}$. Then,\vspace{-0.2cm}
\begin{eqnarray}\label{simpil_u_underline_bef}
&&\hspace{-0.8cm}\sum_{i\in V}\!\sum_{j\in N_i}\!(u_i^{ij}(k+1)\!+\!v_i^{ji}(k+1)\!-\!{u_i^{ij}}^*\!-\!{v_i^{ji}}^*)^T\!(x_i^i(k)\!-\!x_i^*)\nonumber\\
&&\hspace{-0.8cm}+\sum_{i\in V}\sum_{j\in N_i}(u_{-i}^{ij}(k+1)+v_{-i}^{ji}(k+1))^T(x_{-i}^i(k)-x_{-i}^*)\nonumber\\
&&\hspace{-0.8cm}=\sum_{i\in V}\sum_{j\in N_i}(u_i^{ij}(k+1)-{u_i^{ij}}^*)^T(x_i^i(k)-x_i^*)\nonumber\\
&&\hspace{-0.8cm}+\sum_{i\in V}\sum_{j\in N_i}(v_i^{ij}(k+1)-{v_i^{ij}}^*)^T(x_i^j(k)-x_i^*)\nonumber\\
&&\hspace{-0.8cm}+\sum_{i\in V}\sum_{j\in N_i}u_{-i}^{ij}(k+1)^T(x_{-i}^i(k)-x_{-i}^*)\nonumber\\
&&\hspace{-0.8cm}+\sum_{i\in V}\sum_{j\in N_i}v_{-i}^{ij}(k+1)^T(x_{-i}^j(k)-x_{-i}^*).
\end{eqnarray}
Note that by \eqref{u_update_ADMM} and \eqref{v_update_ADMM} as well as the initial conditions for Lagrange multipliers $u^{ij}(0)=v^{ij}(0)=\textbf{0}_N$ $\forall i\in V,\,j\in N_i$, we obtain, \begin{equation}\label{u+v=0}u^{ij}(k)+v^{ij}(k)=\textbf{0}_N\quad \forall i\in V,\,j\in N_i,\,k>0.
\end{equation} 
Substituting \eqref{u+v=0} into \eqref{simpil_u_underline_bef} and using \eqref{PNash2}, we obtain,\vspace{-0.2cm}
\begin{eqnarray}\label{simpil_u_underline}
&&\hspace{-0.8cm}\sum_{i\in V}\sum_{j\in N_i}(u_i^{ij}(k+1)-{u_i^{ij}}^*)^T(x_i^i(k)-x_i^j(k))\nonumber\\
&&\hspace{-0.8cm}+\sum_{i\in V}\sum_{j\in N_i}u_{-i}^{ij}(k+1)^T(x_{-i}^i(k)-x_{-i}^j(k))\nonumber\\
&&\hspace{-0.8cm}=\sum_{i\in V}\sum_{j\in N_i}(u^{ij}(k+1)-{u_i^{ij}}^*e_i)^T(x^i(k)-x^j(k))\nonumber\\
&&\hspace{-0.8cm}=\frac{2}{c}\sum_{i\in V}\sum_{j\in N_i}(u^{ij}(k+1)-{u_i^{ij}}^*e_i)^T(u^{ij}(k+1)-{u^{ij}}(k))\nonumber\\
&&\hspace{-0.8cm}:=\frac{2}{c}(\underline{u}(k+1)-\underline{u}^*)^T(\underline{u}(k+1)-\underline{u}(k)),
\end{eqnarray}
where $\underline{u}=(u^i)_{i\in V}\in \mathbb{R}^{N\sum_{i\in V}|N_i|}$ and $u^i=(u^{ij})_{j\in N_i}\in\mathbb{R}^{N|N_i|}$ and also $\underline{u}^*=({u^i}^*)_{i\in V}\in \mathbb{R}^{N\sum_{i\in V}|N_i|}$ and ${u^i}^*=({u_i^{ij}}^*)_{j\in N_i}\otimes e_i\in\mathbb{R}^{N|N_i|}$.

Using \eqref{Cocoe_in_equat} and \eqref{simpil_u_underline}, for $\rho=\frac{1}{2\sigma_F}$ \eqref{optim_equation_Nash_combo_multiplic_Revised_Sum} becomes,\vspace{-0.3cm}
\begin{eqnarray}\label{optim_equation_Nash_combo_multiplic_Revised_Sum_simpil}
&&\hspace{-0.8cm}-\frac{1}{2}\|\underline{x}(k)-\underline{x}(k-1)\|_{M_1}^2\nonumber\\
&&\hspace{-0.8cm}+(\underline{x}(k)-\underline{x}(k-1))^TM_2(\underline{x}(k)-\underline{x}^*)\nonumber\\
&&\hspace{-0.8cm}+\frac{2}{c}(\underline{u}(k+1)-\underline{u}^*)^T(\underline{u}(k+1)-\underline{u}(k))\leq 0,
\end{eqnarray}
where $M_2:=\text{diag}((\beta_ie_ie_i^T)_{i\in V})+c((D+A)\otimes I_N)$. Note that $\text{diag}((\beta_ie_ie_i^T)_{i\in V})\succeq 0$. Note also that,
\begin{eqnarray}\label{D+A_L}
c((D+A)\otimes I_N)&\!=\!&c((2D-L)\otimes I_N)\nonumber\\
&\!=\!&c((D^{\frac{1}{2}}(2I-D^{-\frac{1}{2}}LD^{-\frac{1}{2}})D^{\frac{1}{2}})\otimes I_N)\nonumber\\
&\!=\!&c((D^{\frac{1}{2}}(2I-L_{N})D^{\frac{1}{2}})\otimes I_N),
\end{eqnarray} 
where $L:=D-A$, $D^{\frac{1}{2}}$, $D^{-\frac{1}{2}}$ and $L_N:=D^{-\frac{1}{2}}LD^{-\frac{1}{2}}$ are the Laplacian of $G_C$, the square root and reciprocal square root of $D$ and the normalized Laplacian of $G_C$, respectively. Since $D\succ 0$, $D^{-\frac{1}{2}}$ exist, it is shown in \cite{chung1997spectral} that $\lambda_{\max}(L_N)\leq2$. Then \eqref{D+A_L} yields that $c((D+A)\otimes I_N)\succeq0$. This concludes $M_2\succeq0$.

We use the following inequality in \eqref{optim_equation_Nash_combo_multiplic_Revised_Sum_simpil} for every $\{a(k)\}$ and $M\succeq 0$:\vspace{-0.3cm}
\begin{eqnarray}
&&\hspace{-1cm}(a(k)-a(k-1))^TM(a(k)-a^*)=\frac{1}{2}\|a(k)-a^*\|_M^2\nonumber\\
&&\hspace{-1cm}+\frac{1}{2}\|a(k)-a(k-1)\|_M^2-\frac{1}{2}\|a(k-1)-a^*\|_M^2.
\end{eqnarray}
Then, \eqref{optim_equation_Nash_combo_multiplic_Revised_Sum_simpil} becomes,\vspace{-0.3cm}
\begin{eqnarray}\label{optim_equation_Nash_combo_multiplic_Revised_Sum_simpil_Q>0}
&&\hspace{-0.7cm}\frac{1}{2}\|\underline{x}(k)-\underline{x}^*\|_{M_2}^2+\frac{1}{c}\|\underline{u}(k+1)-\underline{u}^*\|^2\leq\nonumber\\
&&\hspace{-0.7cm}\frac{1}{2}\|\underline{x}(k-1)-\underline{x}^*\|_{M_2}^2+\frac{1}{c}\|\underline{u}(k)-\underline{u}^*\|^2\\
&&\hspace{-0.7cm}-\frac{1}{2}\|\underline{x}(k)-\underline{x}(k-1)\|_{M_2-M_1}^2-\frac{1}{c}\|\underline{u}(k+1)-\underline{u}(k)\|^2.\nonumber
\end{eqnarray}
By the condition \eqref{condition}, $M_2-M_1\succ0$. Then \eqref{optim_equation_Nash_combo_multiplic_Revised_Sum_simpil_Q>0} implies the following two results:
\begin{enumerate}
	\item $\frac{1}{2}\|\underline{x}(k)-\underline{x}^*\|_{M_2}^2+\frac{1}{c}\|\underline{u}(k+1)-\underline{u}^*\|^2\rightarrow\theta$, for some $\theta\geq0$,
	\vspace{0.3cm}
	\item $\begin{cases}\underline{x}(k)-\underline{x}(k-1)\rightarrow\textbf{0}_{N^2}\\\underline{u}(k+1)-\underline{u}(k)\rightarrow\textbf{0}_{N\sum_{i\in V}|N_i|}\end{cases}$.
\end{enumerate}
Result 1 implies that the sequences $\{x^i(k)\}$ and $\{u^{ij}(k)\}$ (similarly $\{v^{ij}(k)\}$) are bounded and have limit points denoted by $\tilde{x}^i$ and $\tilde{u}^{ij}$ ($\tilde{v}^{ij}$), respectively. Then, we obtain,
\begin{equation}\label{theta}
\theta=\frac{1}{2}\|\underline{\tilde{x}}-\underline{x}^*\|_{M_2}^2+\frac{1}{c}\|\underline{\tilde{u}}-\underline{u}^*\|^2
\end{equation}

Result 2 yields that $\tilde{x}^i=\tilde{x}^j:=\tilde{x}$ for all $i\in V,\,j\in N_i$ since by \eqref{u_update_ADMM} we have,\vspace{-0.3cm}
\begin{eqnarray}\label{xi=xj=tilde}
&&\hspace{-0.8cm}\frac{c}{2}\big(x^i(k)-x^j(k)\big)=u^{ij}(k+1)-u^{ij}(k)\rightarrow\textbf{0}_N\nonumber\\
&&\hspace{-0.8cm}\Rightarrow \tilde{x}^i=\tilde{x}^j\quad\forall i\in V,\,j\in N_i.
\end{eqnarray} 
Moreover, by \eqref{u+v=0} we arrive at,
\begin{equation}\label{u+v=0tilda}\tilde{u}^{ij}+\tilde{v}^{ij}=\textbf{0}_N\quad \forall i\in V,\,j\in N_i.\end{equation}
Result 2 also implies that by \eqref{optim_equation} and \eqref{xi=xj=tilde},
\begin{equation}\label{optim_equation_tilda}
\nabla_iJ_i(\tilde{x})+\partial\mathcal{I}_{\Omega_i}(\tilde{x}_i)+\sum_{j\in N_i}(\tilde{u}_i^{ij}+\tilde{v}_i^{ji})=0.
\end{equation}	
Comparing \eqref{u+v=0tilda}, \eqref{optim_equation_tilda} with \eqref{Nash_equations}, it follows $\forall i\in V,\,j\in N_i$,\vspace{-0.2cm}
\begin{eqnarray}
&&\hspace{-0.8cm}\tilde{x}^i=x^*\quad(\underline{\tilde{x}}=\underline{x^*}),\label{xtilde=xstar}\\
&&\hspace{-0.8cm}\tilde{u}^{ij}={u^{ij}}^*\quad(\underline{\tilde{u}}=\underline{u^*}).\label{utilde=ustar}
\end{eqnarray} 
Using \eqref{xtilde=xstar} and \eqref{utilde=ustar} in \eqref{theta}, it follows that $\theta=0$. Thus, one can conclude from Result~1 that, $\frac{1}{2}\|\underline{x}(k)-\underline{x}^*\|_{M_2}^2+\frac{1}{c}\|\underline{u}(k+1)-\underline{u}^*\|^2\rightarrow 0$ which completes the proof.
$\hfill\blacksquare$
\begin{remark}\label{31insteadAss3}
Assumption~\ref{Lip_assump} is only used in equation \eqref{Cocoe_in_equat}. It is straightforward to verify that \eqref{Cocoe_in_equat} can be satisfied by Assumption~\ref{Lip_assump} for $y=x^*$ (similar to Assumption~4.4 in \cite{frihauf2012nash}).
\end{remark}\vspace{-0.2cm}
\section{implementation of Algorithm}
For the purpose of implementation, we simplify the algorithm to a more compact representation. One may begin with \eqref{x_-i_Update} and \eqref{x_i^i_Mod_ADMM_1} as follows:\vspace{-0.1cm}
\begin{enumerate}
	\item Let $w^i:=\sum_{j\in N_i}u^{ij}+v^{ji}$. Then by \eqref{u_update_ADMM} and \eqref{v_update_ADMM},
	\begin{equation}\label{w^i(k)}
	w^i(k)=w^i(k-1)+c\sum_{j\in N_i}(x^i(k-1)-x^j(k-1)).
	\end{equation}
	\item By replacing $\sum_{j\in N_i}(u_{-i}^{ij}(k)+v_{-i}^{ji}(k))$ with $w_{-i}^i(k)$ and using \eqref{w^i(k)} in \eqref{x_-i_Update}, after a few manipulations we obtain,\vspace{-0.2cm}
	\begin{eqnarray}
	&&\hspace{-1cm}x_{-i}^i(k)=\frac{1}{|N_i|}\sum_{j\in N_i}x_{-i}^j(k-1))-\frac{1}{2c|N_i|}w_{-i}^i(k-1).\nonumber\label{x_-i_Update_simpl}
	\end{eqnarray}
	\item By differentiating of \eqref{x_i^i_Mod_ADMM_1} w.r.t. $x_i^i$ and equating it to $0$, one can verify that $x_i^i$ can be obtained as:\vspace{-0.2cm}
	\begin{eqnarray}
	&&\hspace{-0.8cm}x_i^i(k)\!=\text{arg}\!\min_{x_i^i\in\mathbb{R}}\!\Big\{\!\mathcal{I}_{\Omega_i}\!(x_i^i)\!+\!\frac{\alpha_i}{2}\!\Big\|x_i^i\!-\!\alpha_i^{-1}\!\Big(\!\beta_ix_i^i(k-1)\!-\!w_i^i(k)\nonumber\\
	&&\hspace{-0.8cm}-\nabla_iJ_i(x^i(k-1))+c\sum_{j\in N_i}\big(x_i^i(k-1)+x_i^j(k-1)\big)\Big)\Big\|^2\Big\},\nonumber\label{x_i^i_Mod_ADMM_simplified}
	\end{eqnarray}
	where $\alpha_i=\beta_i+2c|N_i|$. Let $\text{prox}_{g}^{a}[s]:=\text{arg }\min_{x}\{g(x)+\frac{a}{2}\|x-s\|^2\}$ be the proximal operator for the non-smooth function $g$. Note that $\text{prox}_{\mathcal{I}_{\Omega_i}}^{\alpha_i}[s]=T_{\Omega_i}[s]$ where $T_{\Omega_i}:\mathbb{R}\rightarrow\Omega_i$ is an Euclidean projection. Then for each player $i$, $\forall i\in V$ we obtain,\vspace{-0.2cm}
	\begin{eqnarray}
	&&\hspace{-0.3cm}x_i^i(k)=T_{\Omega_i}\Big[\alpha_i^{-1}(\beta_i+c|N_i|)x_i^i(k-1)\nonumber\\
	&&\hspace{-0.3cm}-\alpha_i^{-1}\Big(w_i^i(k)+\nabla_iJ_i(x^i(k-1))-c\sum_{j\in N_i}x_i^j(k-1)\Big)\Big].\label{x_i^i_Mod_ADMM}\nonumber
	\end{eqnarray}
\end{enumerate}
Then the ADMM algorithm is as follows:\vspace{-0.2cm}
\begin{algorithm}
	\caption{ADMM Algorithm for Implementation}
	\label{ADMMalgorithm}
	\begin{algorithmic}[1]
		\State \textbf{initialization} $x^i(0)\in\Omega$, $w^{i}(0)=\textbf{0}_N$ $\forall i\in V$
		\For{$k=1,2,\ldots$ }
		\For{each player $i\in V$ }
		\State players exchange estimates with the neighbors
		\State $w^i(k)=w^i(k-1)+c\sum_{j\in N_i}(x^i(k-1)-x^j(k-1))$
		\State $x_{-i}^i(k)=\frac{\sum_{j\in N_i}x_{-i}^j(k-1))}{|N_i|}-\frac{w_{-i}^i(k-1)}{2c|N_i|}$
		\State $x_i^i(k)=T_{\Omega_i}\Big[\frac{\beta_i+c|N_i|}{\alpha_i}x_i^i(k-1)-\alpha_i^{-1}\Big(w_i^i(k)$ 
		\Statex \hspace{2.3cm}$+\nabla_iJ_i(x^i(k-1))-c\sum_{j\in N_i}x_i^j(k-1)\Big)\Big]$
		\EndFor
		\EndFor
	\end{algorithmic}
\end{algorithm}\vspace{-0.2cm}
\section{Simulation Results}
In this section, we compare our algorithm with the gradient-based one proposed in \cite{salehisadaghiani2016distributed}.\vspace{-0.2cm}

We consider a wireless ad-hoc network (WANET) with 16 nodes and 16 multi-hop communication links as in \cite{salehisadaghiani2016arxiv}. There are 15 users who aim to transfer data from a source node to a destination node via this WANET. Fig.~1~(a) shows the topology of the WANET in which solid lines represent links and dashed lines display paths that assigned to users to transfer data. Each link has a positive capacity that restricts the users' data flow . 
\begin{figure}
	\vspace{-2.25cm}
	\hspace{-2.23cm}
	\centering
	\includegraphics [scale=0.54]{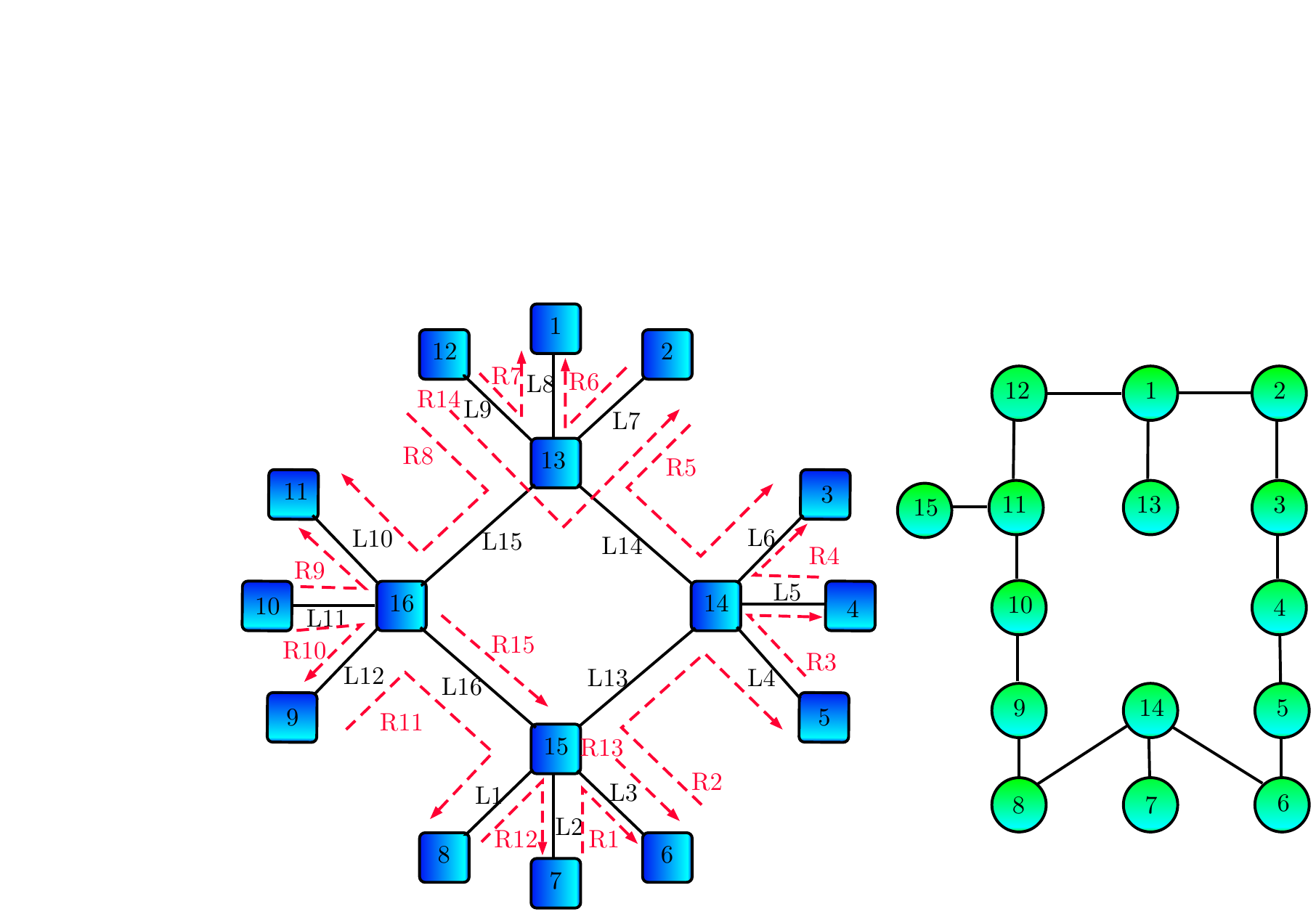}
	\vspace{-0.2cm}
	\caption{(a)  Wireless Ad-Hoc Network (left). (b) Communication graph $G_C$ (right).}
\end{figure}
Here is the list of WANET notations:\vspace{-0.1cm}
\begin{enumerate}
	\item $L_j$: Link $j$, $j\in\{1,\ldots,16\}$,
	\item $R_i$: The path assigned to user $i$, $i\in\{1,\ldots,15\}$,
	\item $C_j>0$: The capacity assigned to each link $j$, $j\in\{1,\ldots,16\}$, 
	\item $0\leq x_i\leq 10$: The data flow of user $i$, $i\in\{1,\ldots,15\}$.
\end{enumerate}
Note that each path consists of a set of links, e.g., $R_1=\{L_2,L_3\}$.

For each user $i$, a cost function $J_i$ is defined as in \cite{salehisadaghiani2016arxiv}:
\begin{equation*}\label{Cost_fcn_gen}
J_i(x_i,x_{-i}):=\sum_{j:L_j\in R_i}\frac{\kappa}{C_j-\sum_{w:L_j\in R_w}x_w}-\chi_i \log(x_i+1),\end{equation*}
where $\kappa>0$ and $\chi_i>0$ are network-wide known and user-specific parameters, respectively.

The problem is to find an NE of the game which is played over a communication graph $G_C$ (depicted in Fig.~1~(b)). It is straightforward to check the Assumptions 1,2 and 3\footnote[4]{It is sufficient for the cost functions that only satisfy equation \eqref{Cocoe_in_equat} (see Remark~\ref{31insteadAss3}).} on $G_C$ and the cost functions. We aim to compare the convergence rate of our algorithm with the one proposed in \cite{salehisadaghiani2016distributed}. The results of Algorithm~1 and the algorithm in \cite{salehisadaghiani2016distributed} are shown in Fig.~2, for $\chi_i=10$ $\forall i\in\{1,\ldots,15\}$ and $C_j=10$ $\forall j\in\{1,\ldots,16\}$ (Fig.~2).
\begin{figure}
	\vspace{-3.98cm}
	\hspace{-0.8cm}
		\includegraphics [scale=0.48]{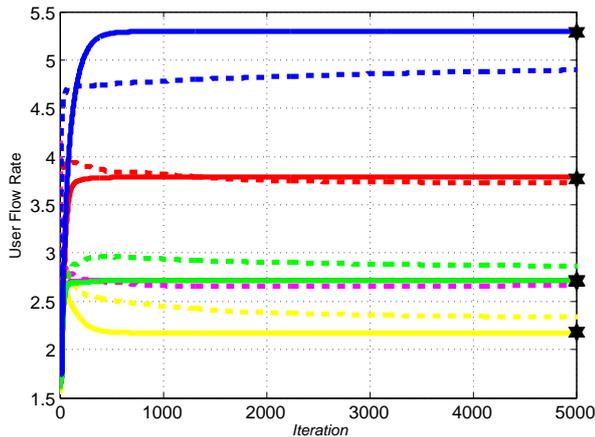}
		\label{fig:minipage1}
	\vspace{-4.4cm}
	\caption{Flow rates of users 1, 3, 5, 8, 13 using our algorithm (solid lines) vs. proposed algorithm in \cite{salehisadaghiani2016distributed} (dashed lines). NE points are represented by black stars.}
\end{figure}
The simulation results verify that the proposed algorithm is 70 times faster than the one in \cite{salehisadaghiani2016distributed}. The factors that lead to this improvement are as follows:\vspace{-0.2cm}
\begin{itemize}
	\item We used the difference between the estimates of the users as a penalty term to update each user's action and estimates.
	\item We used a synchronous algorithm by which every user updates his action and estimates at the same time with the other users.
	\item Unlike gossiping protocol, which is used in \cite{salehisadaghiani2016distributed}, every user communicates with all of the neighboring users (not only one of them) at each iteration.
\end{itemize}\vspace{-0.2cm}
\section{Conclusions}
A distributed NE seeking algorithm is designed using inexact-ADMM to achieve more speed and robustness. The game is reformulated within the framework of inexact-ADMM. The communications between the players are defined to exchange the estimates. An inexact-ADMM-like approach is then designed and its convergence to an NE of the game is analyzed. Eventually, the convergence rate of the algorithm is compared with an existing gossip-based NE seeking algorithm.\vspace{-0.25cm}

\bibliography{ref}
\end{document}